\documentclass[12pt]{elsart}

\usepackage{amsmath}
\usepackage{amsfonts}
\usepackage{graphicx}
\usepackage{amssymb}
\usepackage{ulem}

\begin{document}

\begin{frontmatter}

\title{Statistical measures and the Klein tunneling \\ in single-layer graphene}

\author[jsr]{Jaime Sa\~{n}udo}
\ead{jsr@unex.es} and
\author[rlr]{Ricardo L\'{o}pez-Ruiz}
\ead{rilopez@unizar.es}

\address[jsr]{
Departamento de F\'isica, Facultad de Ciencias, \\
Universidad de Extremadura, E-06071 Badajoz, Spain, \\
and BIFI, Universidad de Zaragoza, E-50009 Zaragoza, Spain}

\address[rlr]{
DIIS and BIFI, Facultad de Ciencias, \\
Universidad de Zaragoza, E-50009 Zaragoza, Spain}

%\date{\today}

\begin{abstract}
Statistical complexity and Fisher-Shannon information are calculated in a \mbox{problem}
of quantum scattering, namely the Klein tunneling across a potential barrier 
in graphene. The treatment of electron wave functions as masless Dirac fermions 
\mbox{allows} us to compute these statistical measures. The comparison of these magnitudes
with the transmission coefficient through the barrier is performed. 
We show that these statistical measures take their minimum values in the situations
of total transparency through the barrier, a phenomenon highly anisotropic for
the Klein \mbox{tunneling} in graphene.  
\end{abstract}


\end{frontmatter}

\maketitle

The calculation of entropic measures in quantum bound states 
\cite{panos2005,sanudo2008-,ferez2009,lopezruiz2011,sanudo2012}
has revealed certain connections with physical properties, 
such as the ionization potential and the static dipole polarizability \cite{sen2007}, 
the closure of shells \cite{panos2009,sanudo2009} 
and the trace of magic numbers \cite{lopezruiz2010,sanudo2011} in atoms and nuclei.

The evaluation of these magnitudes for no bound states can have interest in scattering processes.
A typical scattering process is the crossing of potential barriers by quantum particles \cite{cohen},
a situation which is found in many quantum phenomena, such as tunneling \cite{hartman1962}, 
interferences \cite{perez2001}, resonances \cite{susan1994}, electron transport \cite{zhou2010}, etc.

In a previous work \cite{lopezruiz2013}, the relationship of these entropy-information measures 
with the reflection coefficient in a potential barrier has been investigated. It has been put in evidence 
that these statistical magnitudes present their minimum values
on {\it the transparency points}, just the situations in which the reflection coefficient is null
and then the total transmission through the barrier is achieved.

Following this line, we are also concerned in this work with the calculation of entropic magnitudes 
in the context of scattering processes through potential barriers, in particular with the 
so-called Klein paradox \cite{klein1929}. This is an \mbox{exotic} phenomenon with counterintuitive consequences in 
relativistic quantum mechanics (RQM) that has deserved a great 
interest in particle, nuclear and astro-physics \cite{sakurai1967,greiner1985,su1993,dombey1999,krekora2004,robinson2012}.

From the point of view of the non-relativistic quantum mechanics (NRQM), we know that even if the energy of particles 
in the incident beam is less than the height of the potential barrier, some of the particles can trespass the barrier
(tunneling effect). The higher and wider the potential barrier is, less number of particles
can go through the barrier (exponential decay). Klein \cite{klein1929} showed that, in the frame of RQM, it is possible
that all particles (electrons in this case) can cross the barrier, independently of how high and wide the potential
barrier is. So, in this paradoxical context, it is possible to have situations of total transparency. 
This relativistic effect is explained as a consequence of the matching between electron and positron wavefunctions across 
the barrier, that allows the tunneling of the electrons. The appearance of positrons inside the barrier can be understood
as a result of the sufficiently strong potential of the barrier that is repulsive for the electrons but attractive for the 
positrons (holes) \cite{sakurai1967}.

The Klein paradox has never been realized in laboratory experiments due to the fact that huge electric fields,
$10^{16}$ $Vcm^{-1}$, are necessary for its observation \cite{greiner1985}. However, a possibility to observe this 
type of phenomenon that requires lower electric fields has recently been proposed and tested 
by means of graphene \cite{geim2006,stander2009,young2009}.

Graphene is a new material with a promising potential for many technological applications \cite{novoselov2004,katsnelson2012}. 
It is a monolayer of carbon atoms densely packed in a honeycomb lattice, so it can be considered 
a two-dimensional (2D) system. In this material, electric fields of order $10^5$ $Vcm^{-1}$,
which are routinely created in realistic samples \cite{novoselov2004}, are sufficient to check experimentally 
the Klein tunneling for elementary particles \cite{stander2009,young2009}.

From its electronic properties, graphene is a 2D zero-gap semiconductor with low-energy quasiparticles \cite{geim2006}.
These quantum particles can be viewed as massless relativistic fermions that exhibit, at low Fermi energies
($<1 eV$), a linear dispersion relationship for their energy $E$,

\begin{equation}
E=\hbar k v_{\mbox{\tiny F}},
\end{equation}

where $\hbar$ is the Planck's constant, $\hbar k$ is the quasiparticle momentum with $k$ the wavevector, 
and $v_{\mbox{\tiny F}}$ (instead of $c$ for photons) is the Fermi velocity, $v_{\mbox{\tiny F}}\approx c/300\approx 10^6 ms^{-1}$.

These low-energy quasiparticles can be described by the 2D Dirac-like Hamiltonian given by \cite{geim2006,bai2007}

\begin{equation}
H_0=-i\hbar  v_{\mbox{\tiny F}} \vec{\sigma}\cdot\vec{\nabla},
\end{equation}

where $\vec{\sigma}\equiv (\sigma_x,\sigma_y)$ are the Pauli matrices.

In order to mimic a tunneling experiment in graphene similar to that proposed by Klein, 
a two-dimensional ($x-y$ plane) square potential barrier $V(x)$ is considered:

\begin{equation}
V(x) = \left\{
\begin{array}{rcl}
0, & \mbox{   }x\leq 0 & \mbox{(Region I),} \\
V_0, & \mbox{   }0<x<L & \mbox{(Region II),} \\
0, & \mbox{   }x\geq L  & \mbox{(Region III),} 
\end{array} \right.
\label{eq1}
\end{equation}

where $V_0$ and $L$ are the height and width of the barrier, respectively,
and the dimension along the $Y$ axis is supposed to be infinite. This local potential 
barrier can be created by the electric field effect using local chemical doping or
by means of a thin insulator \cite{novoselov2004,novoselov2005}.
The effect of this potential barrier is to invert charge carriers inside it,
in such a way that holes play the role of positrons, or viceversa.

If an incident electron wave propagates under the action of the Hamiltonian $H=H_0+V(x)$
at an incident angle $\phi$ respect to the $x$ axis, the solutions in the different Regions for 
the two-component Dirac spinor representing these quasiparticles are:

\begin{eqnarray}
\Psi_I(x,y) & = &  \left\{a\left(\begin{array}{c} 1 \\ se^{i\phi}\end{array}\right)e^{ik_xx}
+ r\left(\begin{array}{c} 1 \\ -se^{-i\phi}\end{array}\right)e^{-ik_xx}\right\}\,e^{ik_yy},  \\
\Psi_{II}(x,y) & = &  \left\{A\left(\begin{array}{c} 1 \\ s'e^{i\theta}\end{array}\right)e^{iq_xx}
+ B\left(\begin{array}{c} 1 \\ -s'e^{-i\theta}\end{array}\right)e^{-iq_xx}\right\}\,e^{ik_yy},  \\
\Psi_{III}(x,y) & = &  \;\; t\left(\begin{array}{c} 1 \\ se^{i\phi}\end{array}\right)e^{ik_xx}\,e^{ik_yy},
\end{eqnarray}

where $k_x=k_{\mbox{\tiny F}}\cos\phi$,  $k_y=k_{\mbox{\tiny F}}\sin\phi$ are the wavevector components 
outside the barrier, $k_{\mbox{\tiny F}}=\sqrt{E^2 / \hbar^2 v_{\mbox{\tiny F}}^2}$ is the Fermi wavevector,
$q_x=\sqrt{(E-V_0)^2 / \hbar^2 v_{\mbox{\tiny F}}^2-k_y^2}$ is a wavevector component inside the barrier,
$\theta=\tan^{-1}(k_y/q_x)$ is the refraction angle, $s=sgn(E)$ and $s'=sgn(E-V_0)$.
The five amplitudes ($a, r, A, B, t$) are complex 
numbers determined, up to a global phase factor, by the normalization condition and the boundary 
constraints, namely the continuity of the two components of the Dirac spinor at $x=0$ and $x=L$.

The scattering region (Region II) provokes a partial reflection of the incident electron wave. The reflection
coefficient $R$ gives account of the proportion of the incoming electron flux that is reflected by the barrier.
The expression for $R$ is:
\begin{equation}
R = {Flux_{reflected} \over Flux_{incident}}={|r|^2 \over |a|^2}\,,
\end{equation}
where in this case,
\begin{equation}
{r \over a} = {2e^{i\phi}\left(\sin\phi-ss'\sin\theta\right)\sin(q_xL)
\over ss'\left\{e^{-iq_xL}\cos(\phi+\theta)+e^{iq_xL} \cos(\phi-\theta)\right\}-2i\sin(q_xL)}\,.
\end{equation}

\begin{figure}[h]
\centerline{\includegraphics[width=7cm]{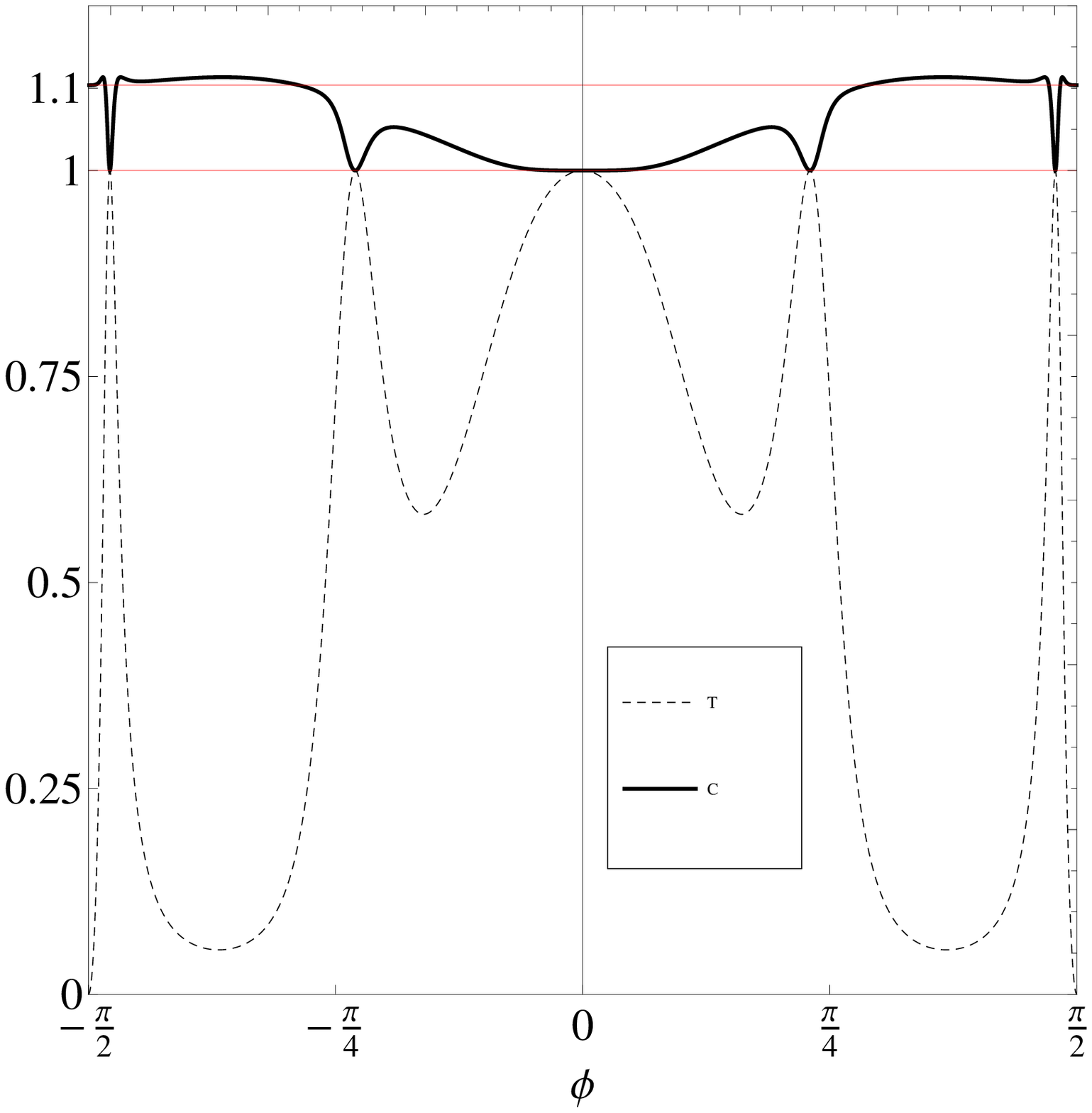}\hskip 5mm\includegraphics[width=7cm]{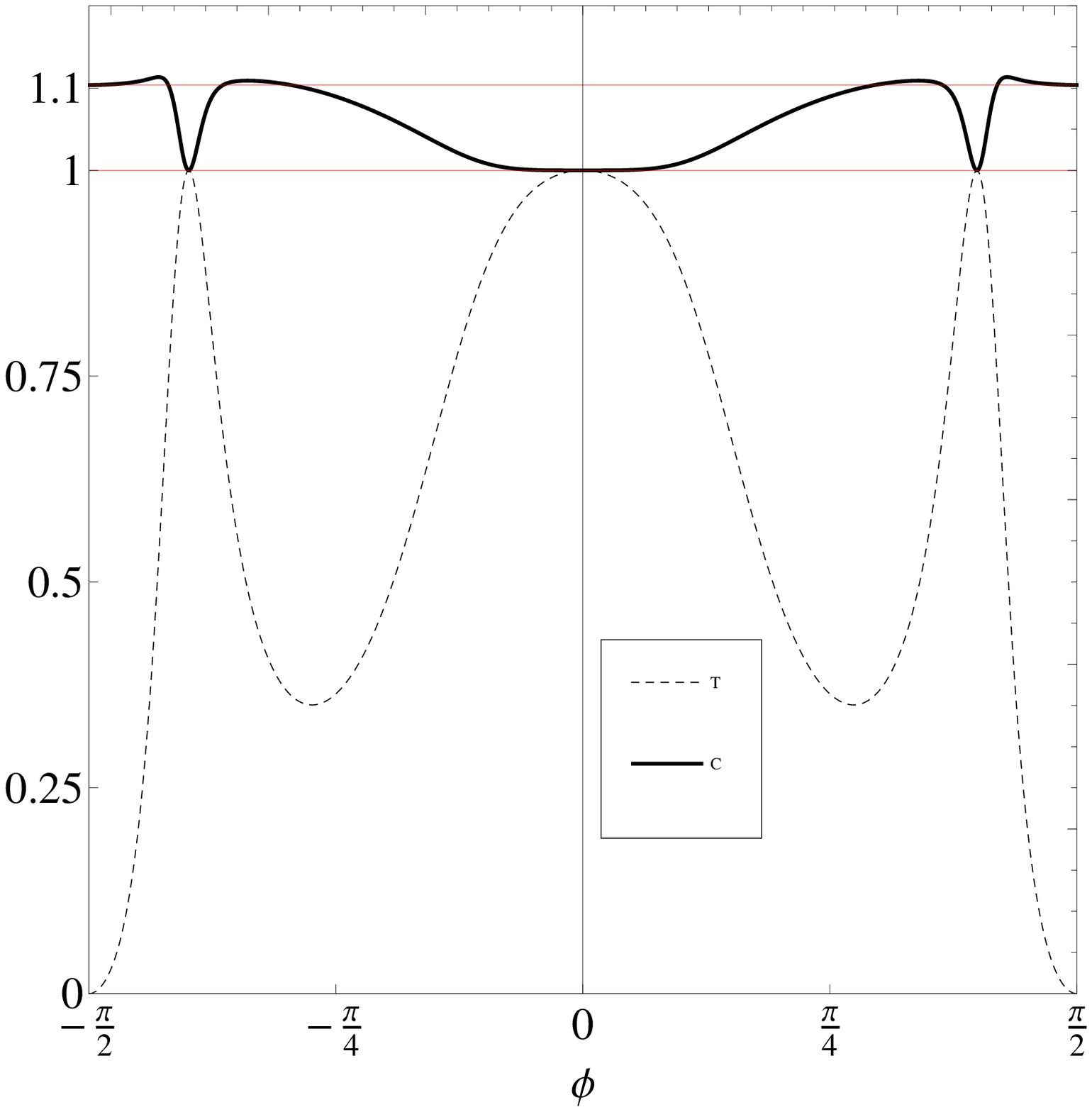}}
\centerline{(a)\hskip 7cm (b)} 
\caption{Statistical complexity, $C$, in Region I and transmission coefficient, $T$, through a 
$100$ nm. wide barrier vs. the incident angle, $\phi$, for single-layer graphene. 
The Fermi energy $E$ of the incident electrons is taken $83$ meV. The barrier heights $V_0$ 
are (a) $200$ and (b) $285$ meV.(Levels $C=1$ and $C=3/e\simeq 1.1$ are plotted in red).}
\label{fig1}
\end{figure}

In this process, there are no sources or sinks of flux, then the transmission coefficient $T$ is given 
by $T=1-R$. This coefficient $T$ is plotted in Fig. \ref{fig1} (dashed lines) for two different heights
$V_0$ of the potential barrier. Observe the anisotropy of its behavior in the sense that
apart of the normal incidence ($\phi=0$), the total transparency, 
$T=1$, can also be found with other angles of incidence. This is just the paradoxical Klein tunneling due to
the fact that the Fermi energy of the incident electrons is much less than the height of the barrier, $E\ll V_0$.
This calculation has been performed by taking for the electron and hole concentration, outside and inside the 
barrier, respectively, the typical values used in experiments with graphene (see Ref. \cite{geim2006}).

Now, we proceed to calculate two statistical magnitudes for this problem, the statistical complexity
and the Fisher-Shannon entropy. These magnitudes are the result of a global calculation done on
the probability density $\rho(x,y)$ given by $\rho(x,y)=\Psi^+(x,y)\Psi(x,y)$, taking into account that
the region of integration must be adequate to impose the normalization condition in the two-component Dirac 
spinor. As a consequence of having a pure plane wave in the $Y$ axis, the density is a constant in this
direction. Then, without loss of generality, if we take a length of unity in the $Y$ direction, 
the density does not depend on $y$ variable, $\rho(x,y)\equiv \rho(x)$. 
The densities obtained for the different Regions are:

\begin{eqnarray}
\rho_I(x) & = &  2|a|^2\left\{1+\left|{r\over a}\right|^2+Re\left\{\left({r^*\over a}\right)(1-e^{2i\phi})
e^{2ik_xx}\right\}\right\}, \label{eq-ro1}\\
\rho_{II}(x) & = &  2|a|^2\left\{\left|{A\over a}\right|^2+\left|{B\over a}\right|^2+
Re\left\{\left({AB^*\over |a|^2}\right)(1-e^{2i\theta})e^{2iq_xx}\right\}\right\}, \label{eq-ro2} \\
\rho_{III}(x) & = & 2|t|^2.  \label{eq-ro3}
\end{eqnarray}

To normalize these densities, the length of the integration interval in the $X$ direction 
is taken to be $[-\pi/k_x,0]$, $[0,L]$ and $[L,L+\pi/k_x]$, for Regions I, II and III, respectively.

The statistical complexity $C$ \cite{lopez1995,lopez2002},
the so-called $LMC$ complexity, is defined as
\begin{equation}
C = H\cdot D\;,
\end{equation}
where $H$ is a function of the Shannon entropy of the system and $D$ tells us how far the distribution
is from the equiprobability. Here, $H$ is calculated according to the 
simple exponential Shannon entropy $S$ \cite{lopez2002,shannon1948,dembo1991}, that has the expression, 
\begin{equation}
H = e^{S}\;,
\end{equation}
with
\begin{equation}
S = -\int \rho(x)\;\log \rho(x)\; dx \;.
\label{eq1-S}
\end{equation}
Some kind of distance to the equiprobability distribution is taken for the disequilibrium $D$,
\cite{lopez1995,lopez2002}, that is,
\begin{equation}
D = \int \rho^2(x)\; dx\;.
\label{eq2-D} 
\end{equation}

\begin{figure}[t]
\centerline{\includegraphics[width=7cm]{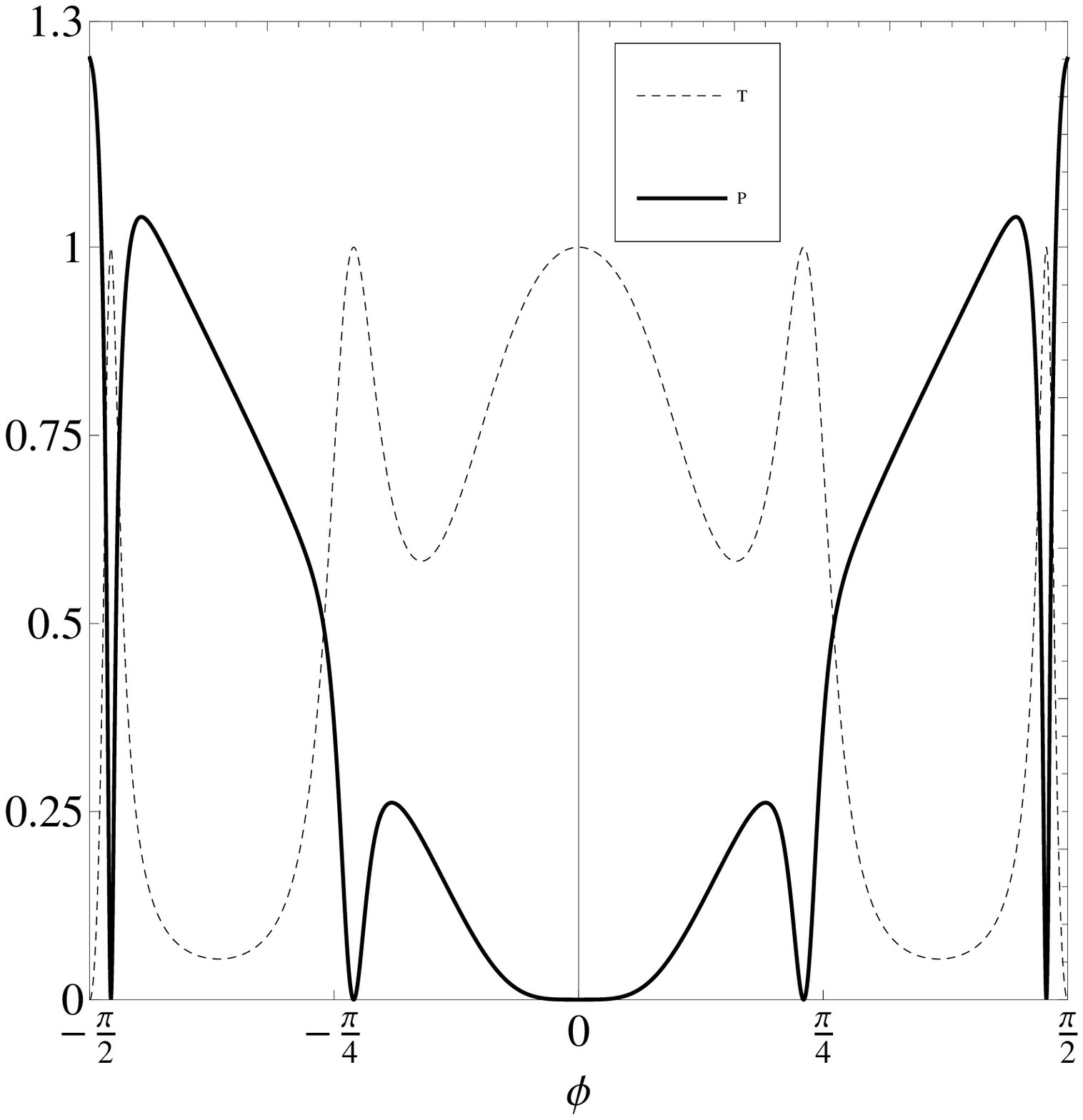}\hskip 5mm\includegraphics[width=7cm]{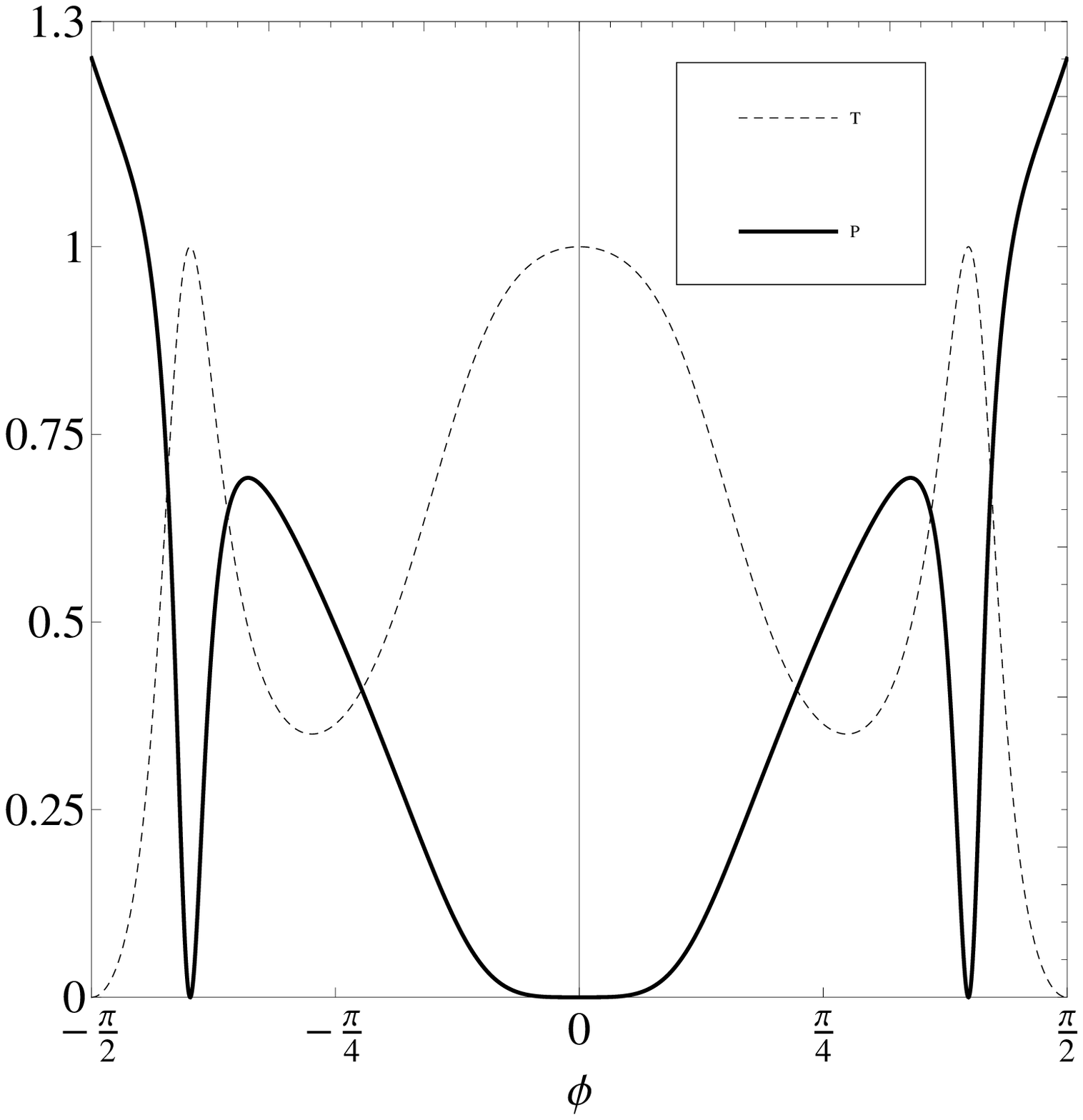}}
\centerline{(a)\hskip 7cm (b)} 
\caption{Fisher-Shannon information, $P$, in Region I and transmission coefficient, $T$, through a 
$100$ nm. wide barrier vs. the incident angle, $\phi$, for single-layer graphene. 
The Fermi energy $E$ of the incident electrons is taken $83$ meV. The barrier heights $V_0$ 
are (a) $200$ and (b) $285$ meV.}
\label{fig2}
\end{figure} 

The Fisher-Shannon information $P$ \cite{romera2004,sen2008} is defined as   
\begin{equation}
P= J\cdot I\,,
\end{equation}
where the first factor is a version of the exponential Shannon entropy \cite{dembo1991}, 
\begin{equation}
J = {1\over 2\pi e}\;e^{2S}\;,
\end{equation}
with the constant $2$ in the exponential selected to have a non-dimensional $P$.
The second factor
\begin{equation}
I = \int {[d\rho(x)/dx]^2\over \rho(x)}\; dx\;,
\end{equation}
is the so-called Fisher information measure \cite{fisher1925}, that quantifies the irregularity
of the probability density.

%%FIGURE1
The statistical complexity $C$ in Region I is plotted in solid line in Fig. \ref{fig1}. 
For the points of total transmission, $T=1$, the Dirac spinor is a pure plane wave in the $X$ direction 
due to the fact that there is not reflected wave, $r=0$ in expression (\ref{eq-ro1}), 
then the density is constant and the complexity $C$ is the minimum, $C=1$. 
For the values of $\phi$ in between the transparency points, there is reflected wave, 
$r\neq 0$ in expression (\ref{eq-ro1}),
that interferes with the incident one giving rise to standing-like waves.   
The complexity of any standing wave is $C=3/e\simeq 1.1036$, a value that also corresponds to
the complexity of the eigenstates of the infinite square well \cite{lopezruiz2009}.
Observe that the points of total transparency are also well signalled by the minima of
the statistical complexity $C$.

%%FIGURE2
In Fig. \ref{fig2}, the behavior of the Fisher-Shannon information $P$ for the Region I 
is plotted in solid line. The transmission coefficient $T$ is also shown
in dashed line. Similarly to the behavior of $C$ in Region I, 
the Dirac spinor is a pure plane wave on the points of total transmission, 
$T=1$, it means that the density is constant, then the Fisher information 
is null and so $P$. In between the transparency points, standing-like waves are formed 
due to interference of the reflected waves with incident ones. The Fisher information
is non null due to the appearance of oscillations in the density, then the Fisher-Shannon 
information takes on these regions the maximum values, presenting the biggest one,
$P=1.2512$, for $\phi=\pm\pi/2$, that is the P value of a perfect standing wave, 
a fact compatible in this case with a total reflection of the electron flux.

\begin{figure}[t]
\centerline{\includegraphics[width=7cm]{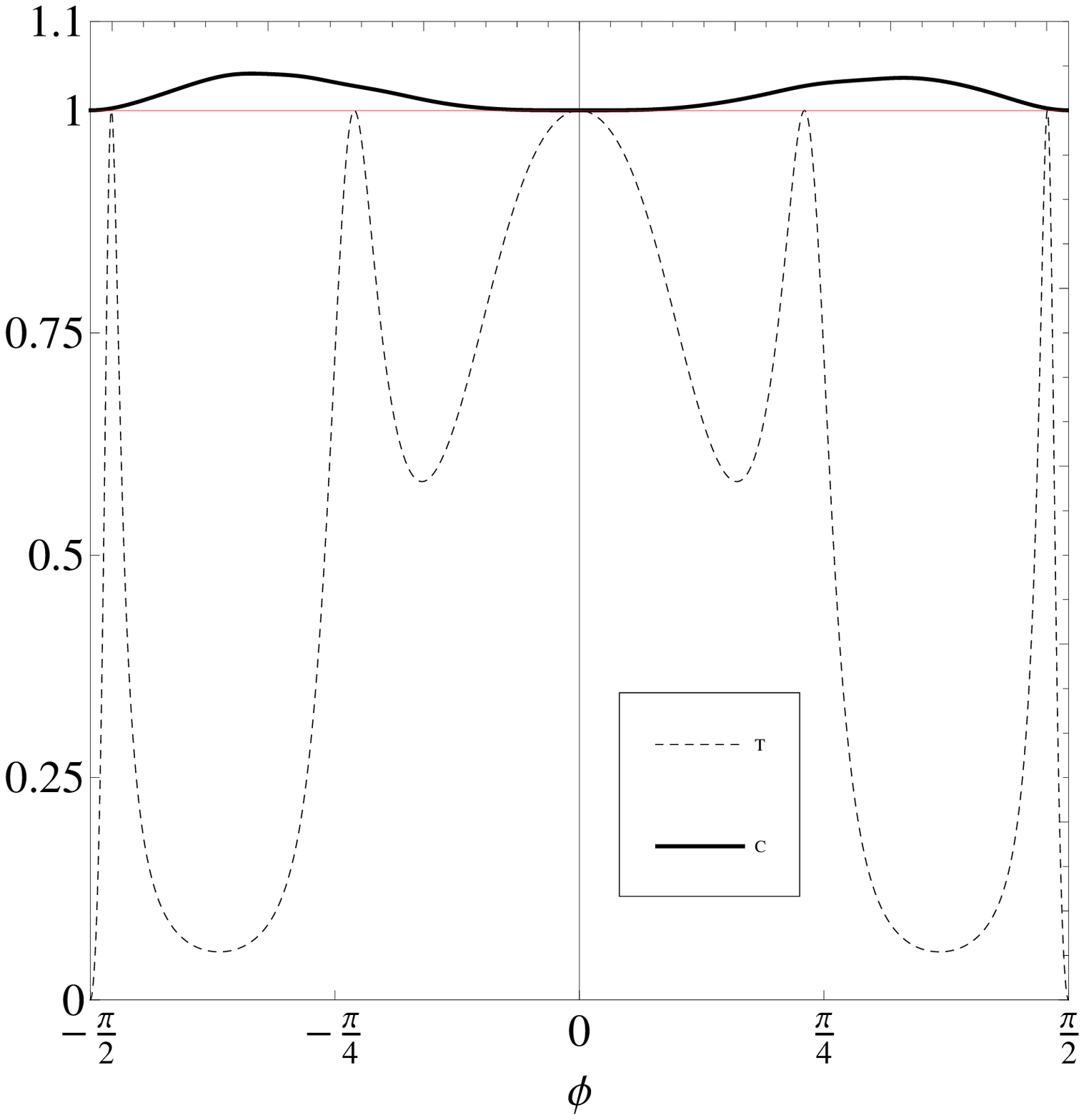}\hskip 5mm\includegraphics[width=7cm]{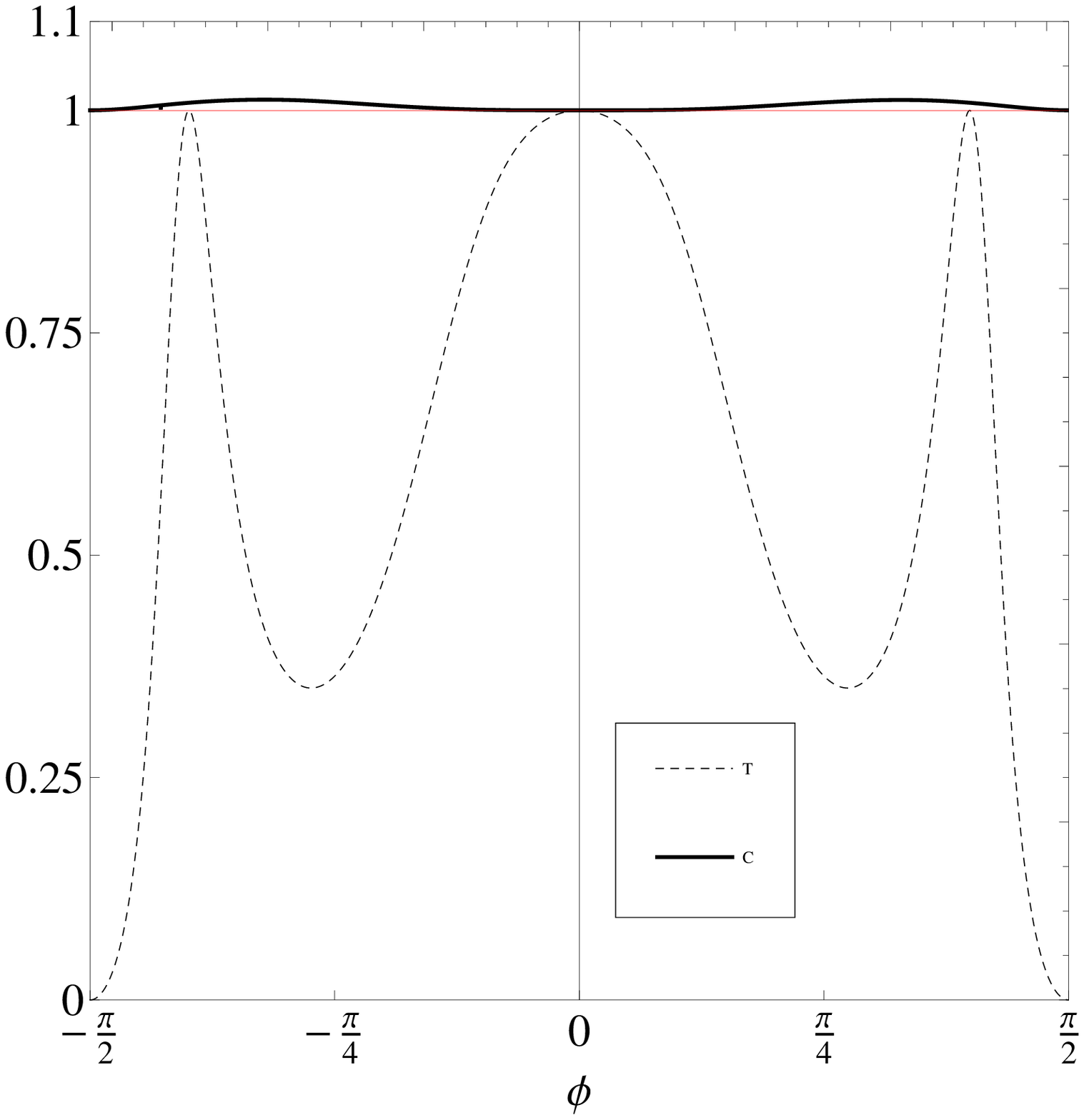}}
\centerline{(a)\hskip 7cm (b)} 
\caption{Statistical complexity, $C$, in Region II and transmission coefficient, $T$, through a 
$100$ nm. wide barrier vs. the incident angle, $\phi$, for single-layer graphene. 
The Fermi energy $E$ of the incident electrons is taken $83$ meV. The barrier heights $V_0$ 
are (a) $200$ and (b) $285$ meV. (Level $C=1$ is plotted in red).}
\label{fig3}
\end{figure}

\begin{figure}[t]
\centerline{\includegraphics[width=7cm]{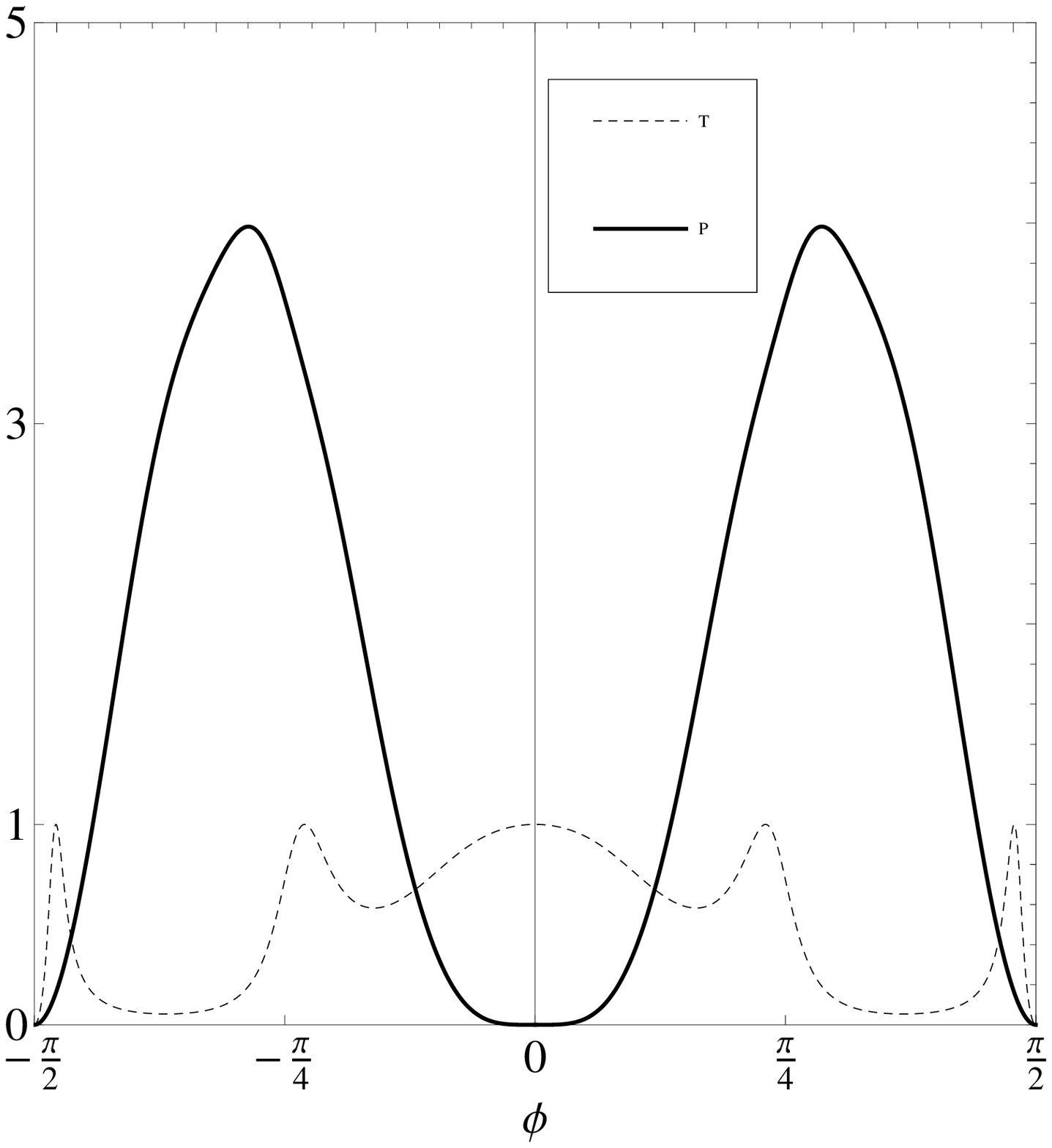}\hskip 5mm\includegraphics[width=7cm]{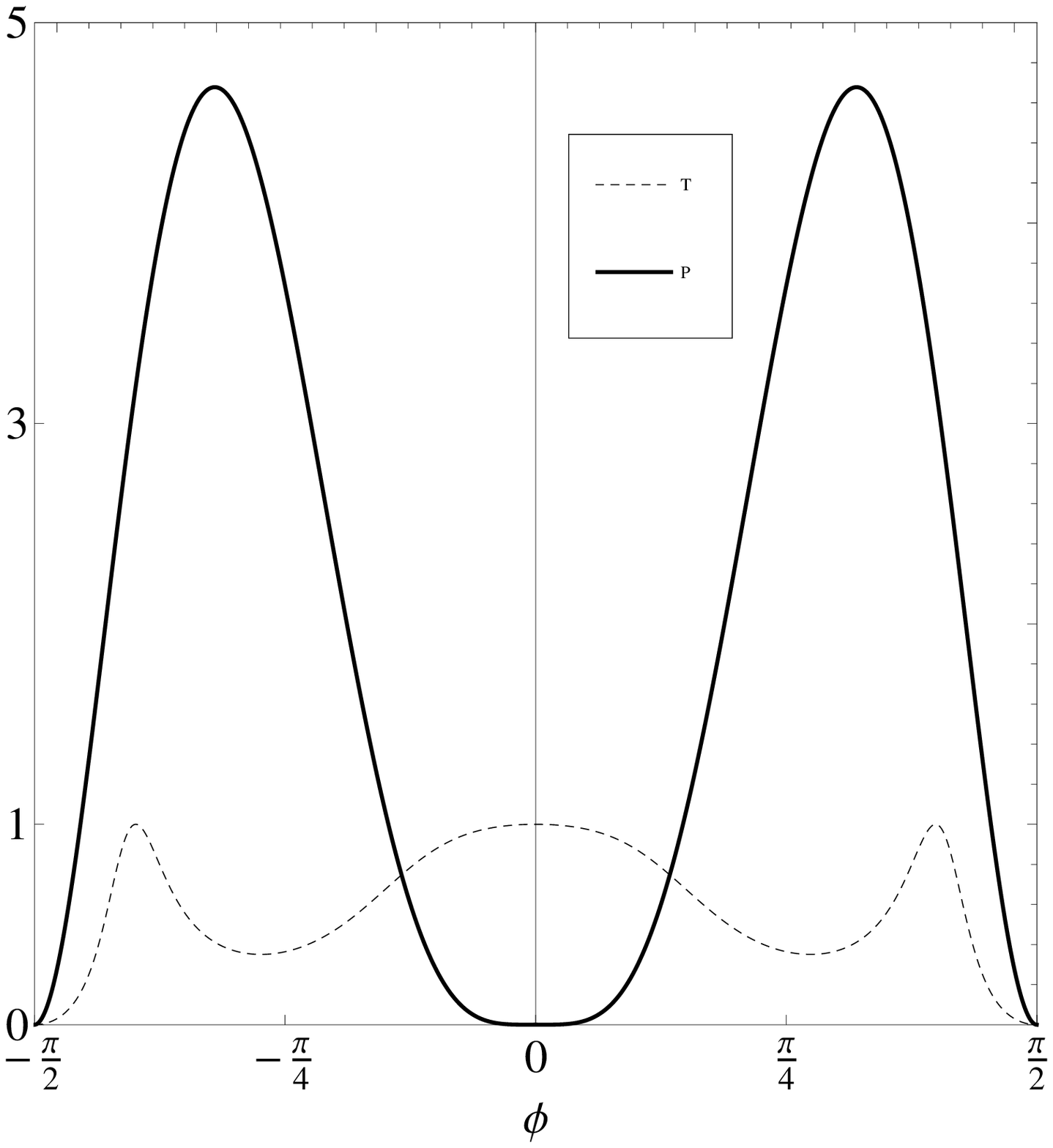}}
\centerline{(a)\hskip 7cm (b)} 
\caption{Fisher-Shannon information, $P$, in Region II and transmission coefficient, $T$, 
through a $100$ nm. wide barrier vs. the incident angle, $\phi$, for single-layer graphene. 
The Fermi energy $E$ of the incident electrons is taken $83$ meV. The barrier heights $V_0$ 
are (a) $200$ and (b) $285$ meV.}
\label{fig4}
\end{figure}

%%FIGURE-3-4
The behavior of $C$ and $P$ for the Region II is plotted in solid line in 
Figs. \ref{fig3} and \ref{fig4}, respectively. The transmission coefficient $T$ is also 
shown in dashed line. When the incidence is normal, $\phi=0$, the density is a constant
as it can be seen in expression (\ref{eq-ro2}), then $C$ and $P$ take their minimum values,
$C=1$ and $P=0$, respectively. For other angles of incidence, the density presents oscillations,
hence $C$ and $P$ take bigger values than their respective minima. For $\phi=\pm\pi/2$, there is
a situation compatible with a total reflection of the electron wave function that means no penetration 
in the barrier, then $C$ and $P$ recover their minimum values.

In the Region III, the density is a constant for any angle of incidence, see expression (\ref{eq-ro3}),
then $C$ and $P$ take their minimum values, $C=1$ and $P=0$.

In conclusion, the calculation of the statistical complexity and the Fisher-Shannon information
in an applied problem of electron scattering on a potential barrier in graphene 
has been presented. The relationship of these indicators with a physical 
magnitude, the transmission coefficient, has been disclosed. When the transmission
through the barrier is complete, these statistical magnitudes take their minimum values,
a fact that put in evidence that these entropic measures can also be useful to detect
certain physical phenomena in quantum problems, the Klein tunneling in graphene included.

\section*{Acknowledgements}
 The authors acknowledge some financial support from  the Spanish project
 DGICYT-FIS2009-13364-C02-01. J.S. also thanks to the Consejer\'ia de 
 Econom\'ia, Comercio e Innovaci\'on of the Junta de Extremadura (Spain) for 
 financial support, Project Ref. GRU09011.

\end{document}